\documentclass[
    ,final            
  ,numberedheadings 
  ]
  {aipproc}

\layoutstyle{6x9}
\usepackage{epsfig}
\usepackage{amsmath}

\begin{document}
\title{Distribution of winners in truel games}

\author{R. Toral and P. Amengual}{
  address={Instituto Mediterr\'aneo de Estudios Avanzados (IMEDEA) CSIC-UIB\\
  Ed. Mateu Orfila, Campus UIB\\
  E-07122 Palma de Mallorca\\
  SPAIN}
}
\begin{abstract}
In this work we present a detailed analysis using the Markov chain
theory of some versions of the truel game in which three players
try to eliminate each other in a series of one-to-one competitions, using the rules of the game. Besides reproducing some known expressions for the winning
probability of each player, including the equilibrium points, we
give expressions for the actual distribution of winners in a truel
competition.
\end{abstract}

\maketitle


\section{Introduction}

A truel is a game in which three players aim to eliminate each
other in a series of one-to-one competitions. The mechanics of the
game is as follows: at each time step, one of the players is
chosen and he decides who will be his target. He then aims at this
person and with a given probability he might achieve the goal of
eliminating him from the game (this is usually expressed as the
players ``shooting" and ``killing" each other, although  possible
applications of this simple game do not need to be so violent).
Whatever the result, a new player is chosen amongst the survivors
and the process repeats until only one of the three players
remains. The paradox is that the player that has the highest
probability of annihilating competitors does not need to be
necessarily the winner of this game. This surprising result was
already present in the early literature on truels, see the
bibliography in the excellent review of reference \cite{KB97}.
According to this reference, the first mention of truels was in
the compendium of mathematical puzzles by Kinnaird \cite{kin46}
although the name {\sl truel} was coined by Shubik \cite{shu82} in
the 1960s.

Different versions of the truels vary in the way the players are
chosen (randomly, in fixed sequence, or simultaneous shooting),
whether they are allowed to ``pass", i.e. missing the shoot on
purpose (``shooting into the air"), the number of tries (or
``bullets") available for each player, etc. The strategy of each
player consists in choosing the appropriate target when it is his
turn to shoot. Rational players will use the strategy that
maximizes their own probability of winning and hence they will
chose the strategy given by the equilibrium Nash point. In a
series of seminal papers\cite{k72.1,k75.1,k75.3}, Kilgour has
analyzed the games and determined the equilibrium points under a
variety of conditions.

In this paper, we analyze the games from the point of view of
Markov chain theory. Besides being able to reproduce some of the
results by Kilgour, we obtain the probability distribution for the
winners of the games. We restrict our study to the case in which
there is an infinite number of bullets and consider two different
versions of the truel: random and fixed sequential choosing of the
shooting player. These two cases are presented in sections
\ref{sec:random} and \ref{sec:seq}, respectively. In section
\ref{sec:opi} we consider a variation of the game in which, instead
of eliminating the competitors from the game, the objective is to
convince them on a topic, making the truel suitable for a model of
opinion formation. Some conclusions and directions for future work
are presented in section \ref{sec:conclusions} whereas some of the
most technical parts of our work are left for the final
appendixes.

\section{Random firing}
\label{sec:random}

Let us first fix the notation. The three players are labeled as
A,B,C. We denote by $a$, $b$ and $c$, respectively, their {\sl
marksmanship}, defined as the probability that a player has of
eliminating from the game the player he has aimed at. The {\sl
strategy} of a player is the set of probabilities he uses in order
to aim to a particular player or to shoot into the air. Obviously,
when only two players remain, the only meaningful strategy is to
shoot at the other player. If three players are still active, we
denote by $P_{AB}$, $P_{AC}$ and $P_{A0}$ the probability of
player A shooting into player B, C, or into the air, respectively,
with equivalent definitions for players B and C. These
probabilities verify $P_{AB}+P_{AC}+P_{C0}=1$. A ``pure" strategy
for player A corresponds to the case where one of these three
probabilities is taken equal to 1 and the other two equal to 0,
whereas a ``mixed" strategy takes two or more of these
probabilities strictly greater than 0. Finally, we denote by
$\pi(a;b,c)$ the probability that the player with marksmanship $a$
wins the game when he plays against two players of marksmanship
$b$ and $c$. The definition implies $\pi(a;b,c)=\pi(a;c,b)$ and
$\pi(a;b,c)+\pi(b;a,c)+\pi(c;a,b)=1$.

In the particular case considered in this section, at each time
step one of the players is chosen {\bf randomly} with equal
probability amongst the survivors. There are 7 possible states of
this system labeled as ABC, AB, AC, BC, A, B, C, according to the
players who remain in the game. The game can be thought of as a
Markov chain with seven states, three of them being absorbent
states. The details of the calculation for the winning
probabilities of A, B and C as well as a diagram of the allowed
transitions between states are left for the
appendix~\ref{ap:random}. We now discuss the results in different
cases.

Imagine that the players do not adopt any thought strategy and
each one shoots randomly to any of the other two players. Clearly,
this is equivalent to setting
$P_{AB}=P_{AC}=P_{BA}=P_{BC}=P_{CA}=P_{CB}=1/2$. The winning
probabilities in this case are:
\begin{equation}
\label{eq:random} \pi(a;b,c) = \frac{a}{a + b + c},~~~~ \pi(b;a,c)
= \frac{b}{a + b + c},~~~~ \pi(c;a,b) =\frac{c}{a + b + c},
\end{equation}
a logical result that indicates that the player with the higher
marksmanship possesses the higher probability of winning. Identical
result is obtained if the players include shooting in the air as
one of their equally likely possibilities.

It is conceivable, though, that players will not decide the
targets randomly, but will use some strategy in order to maximize
their winning probability. Completely rational players will
choose strategies that are best responses (i.e. strategies that
are utility--maximizing) to the strategies used by the other
players. This defines an equilibrium point when all the players
are better off keeping their actual strategy than changing to
another one. Accordingly, this equilibrium point can be defined as
the set of probabilities $P_{\alpha \beta}$ (with
$\alpha=$A,B,C and $\beta=$A,B,C,0) such that the winning probabilities have a
maximum. This set can be found from the expressions
in the appendix, with the result that the equilibrium point in the
case $a>b>c$ is given by $P_{AB}=P_{CA}=P_{BA}=1$ and
$P_{AC}=P_{A0}=P_{BC}=P_{B0}=P_{CB}=P_{C0}=0$. This is the
``strongest opponent strategy'' in which each player aims at the
strongest of his opponents\cite{KB97}. With this strategy, the
winning probabilities are:

\begin{equation}
\label{eq:pirandom} \pi(a;b,c)= \frac{a^2}{(a+c)(a+b+c)},~
\pi(b;a,c) = \frac{b}{a+b+c},~
\pi(c;a,b)=\frac{c(c+2a)}{(a+c)(a+b+c)}
\end{equation}
(notice that these expressions assume $a>b>c$; other cases can be
easily obtained by a convenient redefinition of $a$, $b$ and $c$).

\begin{figure}\label{fig:krandom}
\centerline{\epsfig{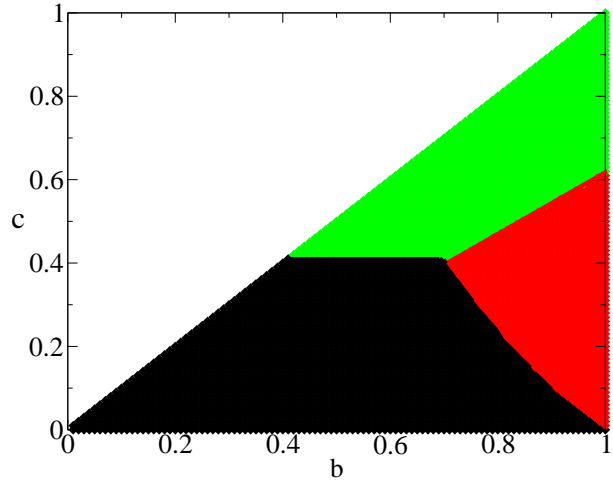}} \caption{In the
parameter space $(b,c)$ with $c<b<a=1$, we indicate by black
(resp. dark gray, light gray) the regions in which player A (resp.
B, C) has the largest probability of winning the truel in the case
of random selection of the shooting player and the use of the
optimal strategy, as given by Eq. (\ref{eq:pirandom}). }
\end{figure}

An analysis of these probabilities leads to the paradoxical result
that when all players use their 'best' strategy, the player with
the worst marksmanship can become the player with the highest
winning probability. For example, if $a=1.0$, $b=0.8$, $c=0.5$
the probabilities of A, B and C winning the game are $0.290$,
$0.348$ and $0.362$, respectively, precisely in inverse order of
their marksmanship. The paradox is explained when one realizes
that all players set as primary target either players $A$ or $B$,
leaving player $C$ as the last option and so he might have the
largest winning probability. In Fig.\ref{fig:krandom} we plot the
regions in parameter space $(b,c)$ (after setting $a=1$)
representing the player with the highest winning probability.

\begin{figure}\label{fig:rrandom}
\centerline{\epsfig{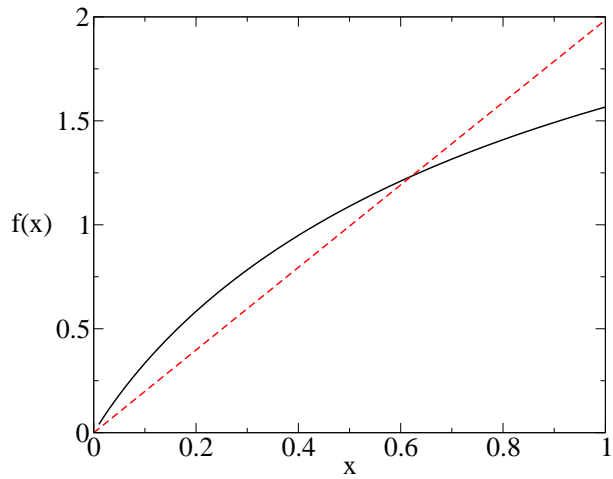}}
\caption{Distribution function $f(x)$ for the winners of truels of
randomly chosen triplets (solid line) in the case of players using
random strategies, Eq. (\ref{eq:random}); distribution $\bar f(x)$
of winners in the case where the winner of a truel remains in the
competition (dashed line).}
\end{figure}

Imagine that we set up a truel competition. Sets of three players
are chosen randomly amongst a population whose marksmanship are
uniformly distributed in the interval $(0,1)$. The distribution of
winners is characterized by a probability density function,
$f(x)$, such that $f(x)dx$ is the proportion of winners whose
marksmanship lies in the interval $(x,x+dx)$. This distribution is
obtained as:
\begin{equation}
f(x)=\int da\,db\,dc\,
\left[\pi(a;b,c)\delta(x-a)+\pi(b;a,c)\delta(x-b)+\pi(c;a,b)\delta(x-c)\right]
\end{equation}
or
\begin{equation}
f(x)= 3\int_0^1 db\int_0^1 dc\,\pi(x;b,c)
\end{equation}
If players use the random strategy, Eq. (\ref{eq:random}), the
distribution of winners is $f(x)=3x\left[x \ln
x-2(1+x)\ln(1+x)+(2+x)\ln(2+x)\right]$. In figure
\ref{fig:rrandom} we observe that, as expected, the function
$f(x)$ attains its maximum at $x=1$ indicating that the best
marksmanship players are the ones which win in more occasions.

We consider now a variation of the competition in which the winner
of one game keeps on playing against other two randomly chosen
players. The resulting distribution of players, $\bar f(x)$, can
be computed as the steady state solution of the recursion
equation:
\begin{equation}
\bar f(x,t+1)=\int da\, db\, dc\,
\left[\pi(a;b,c)\delta(x-a)+\pi(b;a,c)\delta(x-b)+\pi(c;a,b)\delta(x-c)\right]
\bar f(a,t)
\end{equation}
 or
\begin{equation}
\label{eq:recursion} \bar f(x)=\frac{1}{3}\bar f(x)
f(x)+2\int_0^1db\int_0^1dc\,\pi(x;b,c)\bar f(b)
\end{equation}
In the case of using the probabilities of Eq. (\ref{eq:random}) the
distribution of winners is\footnote{The result is more general: if
$\pi(a;b,c)=G(a)/[G(a)+G(b)+G(c)]$, for an arbitrary function
$G(x)$, the solution is $\bar f(x)=G(x)/\int_0^1 G(y)dy$.} $\bar
f(x)=2x$.

\begin{figure}\label{fig:frandom}
\centerline{\epsfig{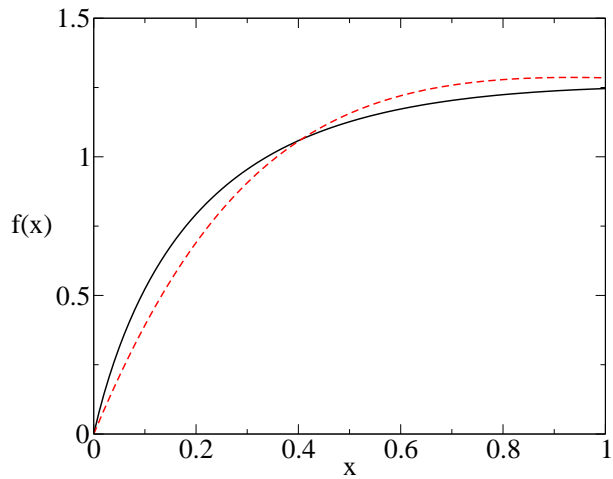}}
\caption{Similar to Fig.(\ref{fig:rrandom}) in the case of the
competition where players use the rational strategy of the
equilibrium point given by eq.(\ref{eq:pirandom}).}
\end{figure}

For players adopting the equilibrium point strategy,
Eq.(\ref{eq:pirandom}), the resulting expression for $f(x)$ is too
ugly to be reproduced here, but the result has been plotted in
Fig. \ref{fig:frandom}. Notice that, despite the paradoxical
result mentioned before, the distribution of winners still has it
maximum at $x=1$, indicating that the best marksmanship players
are nevertheless the ones who win in more occasions. In the same figure, we
have also plotted the distribution $\bar f(x)$ of the competition
in which the winner of a game keeps on playing. In this case, the
integral relation Eq.(\ref{eq:recursion}) has been solved
numerically.

\section{Sequential firing}
\label{sec:seq}

In this version of the truel there is an established order of
firing. The players will shoot in increasing value of their
marksmanship.  i.e. if $a>b>c$ the first player to shoot will be
player $C$, followed by player $B$ and the last to shoot is player
$A$. The sequence repeats until only one player remains. Again,
we have left for the appendix \ref{ap:seq} the details of the
calculation of the winning probabilities. Our analysis of the
optimal strategies reproduces that obtained by the detailed study
of Kilgour\cite{k75.1}. The result is that there are two
equilibrium points depending on the value of the function
$g(a,b,c)= a^2(1-b)^2(1-c)-b^2c - a\,b\,(1-b\,c)$: if $g(a,b,c)>0$
the equilibrium point is the strongest opponent strategy
$P_{AB}=P_{BA}=P_{CA}=1$, while for $g(a,b,c)<0$ it turns out that
the equilibrium point strategy is $P_{AB}=P_{BA}=P_{C0}=1$ where
the worst player C is better off by shooting into the air and
hoping that the second best player B succeeds in eliminating the
best player A from the game.

\begin{figure}
\label{fig:kseq}
\centerline{\epsfig{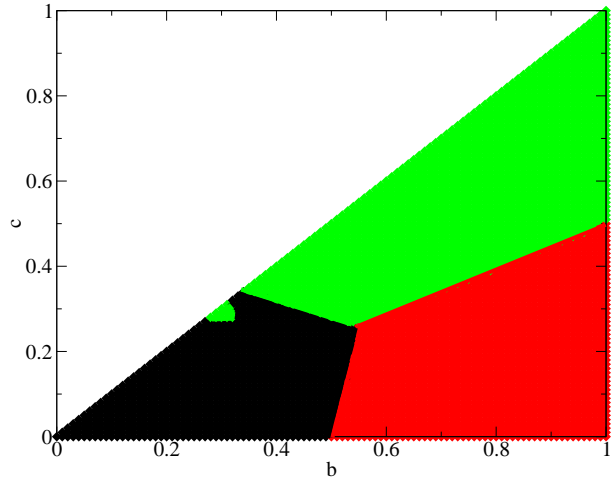}}
\caption{Same as Fig.\ref{fig:krandom} in the case that players
play sequentially in increasing order of their marksmanship. }
\end{figure}

The winning probabilities for this case, assuming $a>b>c$, are:
\begin{eqnarray}
\pi(a;b,c) &=& \frac{(1-c)(1-b)a^2}{[c(1-a)+a][b(1-a)+a]},\nonumber\\
\pi(b;a,c) &=&\frac{(1-c)b^2}{(c(1-b)+b)(b(1-a)+a)} ,\nonumber\\
\pi(c;a,b) &=&\frac{c[b c+a[b (2 + b (-1 + c)-3
c)+c]]}{[c+a(1-c)][b+a(1-b)][a+b(1-a)]},\label{eq:piseq}
\end{eqnarray}
if $g(a,b,c)>0$, and
\begin{eqnarray}
\pi(a;b,c)&=&\frac{a^2 (1-b) (1-c)^2}{[a+(1-a) c] [a+b(1-a)+c(1-a)
(1-b) ]}
,\nonumber\\
\pi(b;a,c)&=&\frac{b \left(b (1-c)^2+c\right)}{[b+
(1-b)c][a+b(1-a)+c (1-a) (1-b)]}
 ,\nonumber\\
\pi(c;a,b)&=&\frac{\frac{a c (1-b) (1-c)}{a+c (1-a)}+\frac{c
(b+c(1-2b))}{b+c(1-b)}}{[a + b (1-a) + c (1-a) (1-b)]
},\label{eq:piseq2}
\end{eqnarray}
if $g(a,b,c)<0$. Again, as in the case of random firing, the
paradoxical result appears that the player with the smallest
marksmanship has the largest probability to win the game. In figure \ref{fig:kseq} we
summarize the results indicating the regions in parameter space
$(b,c)$ (with $a=1$) where each player has the highest probability
of winning. Notice that the 'best' player A has a much smaller
region of winning than compared with the case of random firing.

\begin{figure}\label{fig:fseq}
\centerline{\epsfig{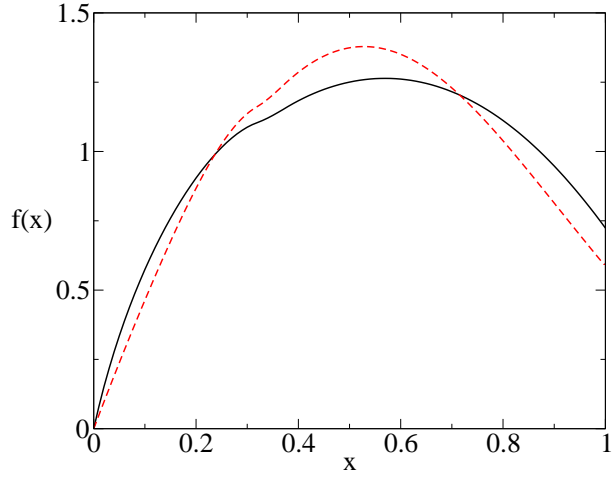}}
\caption{Same as Fig.\ref{fig:rrandom} in the case that players
play sequentially in increasing order of their marksmanship.
Notice that now both distributions of winners present maxima for
$x<1$ indicating that the best a priori players do not win the
game in the majority of the cases.}
\end{figure}

In figure \ref{fig:fseq} we plot the distribution of winners
$f(x)$ and $\bar f(x)$ in a competition as defined in the previous
section. Notice that now the distribution of winners $f(x)$ has a
maximum at $x\approx 0.57$ indicating that the players with the
best marksmanship do not win in the majority of cases.

\section{Convincing opinion }
\label{sec:opi} 
\begin{figure}
\centerline{\epsfig{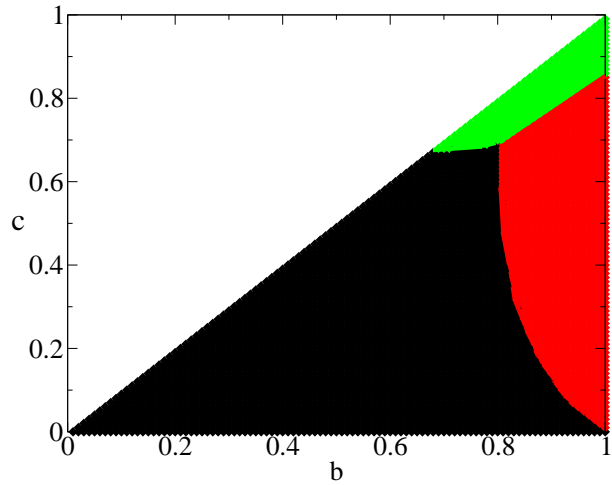}}
\caption{\label{fig:diagram_opinion}Same as Fig.\ref{fig:krandom}
for the convincing opinion model. }
\end{figure}

We reinterpret the truel as a game in which three
people holding different opinions, A, B and C, on a topic, aim to
convince each other in a series of one-to-one discussions. The
marksmanship $a$ (resp. $b$, $c$) are now interpreted as the
probabilities that player holding opinion A (resp. B or C) have of
convincing  another player of adopting this opinion. The main
difference with the previous sections is that now there are always three players present in the
game and the different states in the Markov chain are ABC, AAB,
ABB, AAC, ACC, BBC, BCC, AAA, BBB and CCC.  The analysis of the
transition probabilities is left for appendix \ref{ap:con}. We
consider only the random case in which the person that tries to
convince another one is chosen randomly amongst the three players.
The equilibrium point corresponds to the best opponent strategy
set of probabilities in which each player tries to convince the
opponent with the highest marksmanship. The
probabilities that the final consensus opinion is A, B or C,
assuming $a>b>c$ are given by
\begin{eqnarray}
\pi(a;b,c)&=&\frac{a^2 \left[2 c b^2+a \left((a+b)^2+2 (a+2 b) c\right)\right]}{(a+b)^2 (a+c)^2 (a+b+c)},\nonumber\\
\pi(b;a,c)&=& \frac{b^2(b+3c)}{(b+c)^2(1+b+c)},\nonumber\\
\pi(c;a,b)&=&\frac{c^2 \left[c^3+3 (a+b) c^2+a (a+8 b) c+a b (3
a+b)\right]}{(a+c)^2 (b+c)^2 (a+b+c)} ,\label{opinion3}
\end{eqnarray}
respectively. As shown in Fig.~\ref{fig:diagram_opinion}, there is
still a set of parameter values $(a,b,c)$ for which opinion C has
the highest winning probability, although it is smaller than in
the versions considered in the previous sections.

\begin{figure}\label{fig:fopi}
\centerline{\epsfig{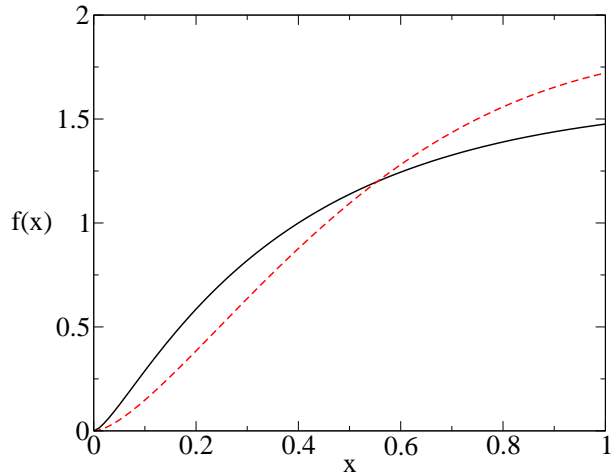}} \caption{Same
as Fig.\ref{fig:rrandom} for the convincing opinion model.}
\end{figure}

Similarly to other versions, we plot in figure \ref{fig:fopi} the
distribution of winning opinions, $f(x)$. Notice that, as in the
random firing case, it attains its maximum at $x=1$ showing that
the most convincing players win the game in more occasions. We
have also plotted in the same figure, the distribution $\bar f(x)$
which results where one of the winners of a truel is kept to
discuss with two randomly chosen players in the next round.

\section{Conclusions}
\label{sec:conclusions} As discussed in the review of reference
\cite{KB97}, truels are of its interest in many areas of social
and biological sciences. In this work, we have presented a
detailed analysis of the truels using the methods of Markov chain
theory. We are able to reproduce in a language which is more
familiar to the Physics community most of the results of the
alternative analysis by Kilgour\cite{k75.1}. Besides computing the
optimal rational strategy, we have focused on computing the
distribution of winners in a truel competition. We have shown that
in the random case, the distribution of winners still has its
maximum at the highest possible marksmanship, $x=1$, despite the
fact that sometimes players with a lower marksmanship have a
higher probability of winning the game. In the sequential firing
case, the paradox is more present since even the distribution of
winners has a maximum at $x<1$. It would be interesting to
determine mechanisms by which players could, in an evolutionary
scheme, adapt themselves to the optimal values.


\section{Appendix: Calculation of the probabilities}
\subsection{Random firing}
\label{ap:random} 
\begin{figure}[!htb]
\begin{tabular}{|c|c|} \hline
\textit{States} & \textit{Remaining players}\\
\hline
$0$ & ABC\\
$1$ & AB\\
$2$ & AC\\
$3$ & BC\\
$4$ & A\\
$5$ & B\\
$6$ & C\\
\hline
\end{tabular}
\epsfig{figure=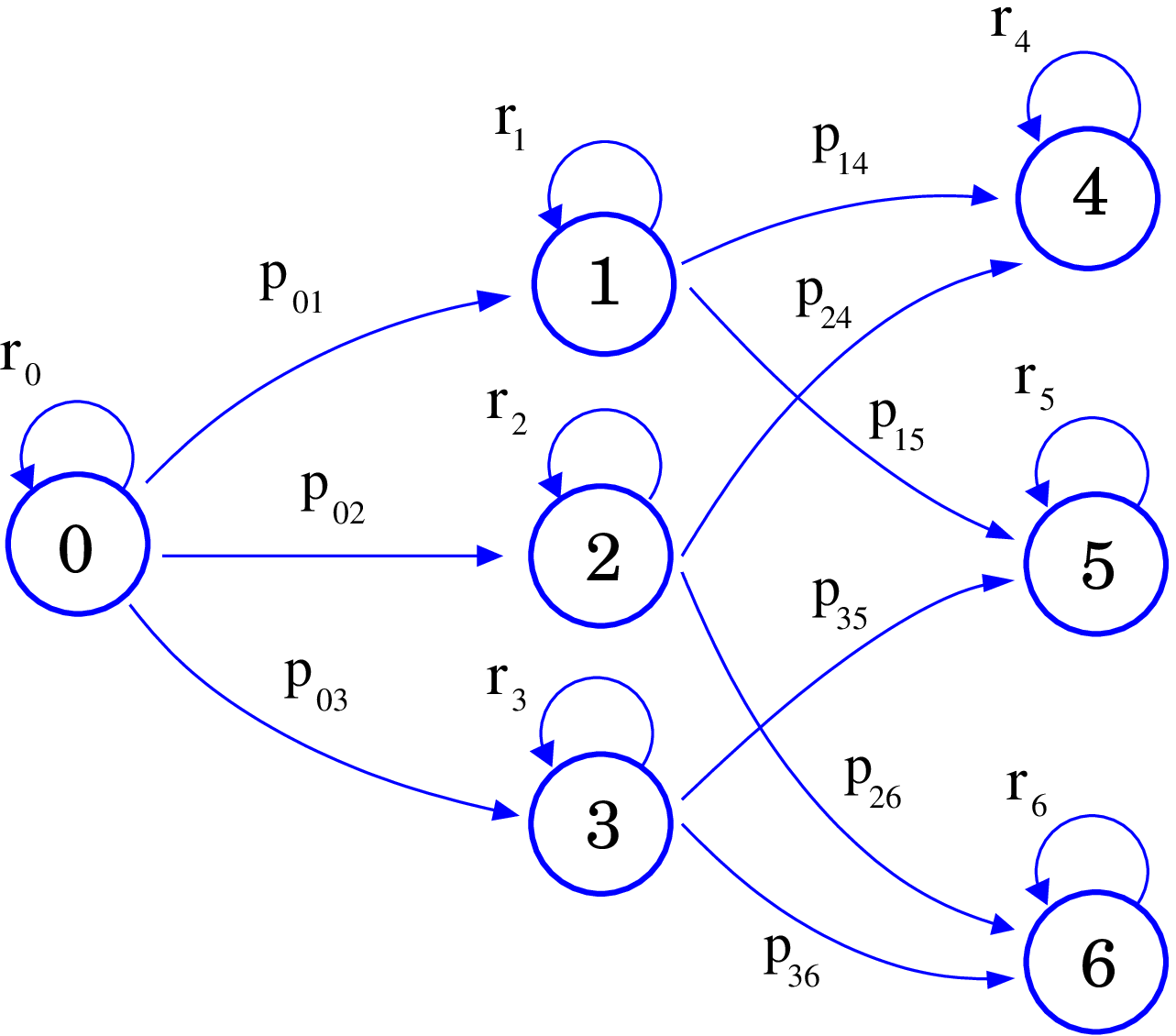,width=7.cm}
\caption{\label{fig:random}Table with the description of all the
possible states for the random firing game, and diagram
representing the allowed transitions between the states shown in
the table. }
\end{figure}

In this game there are seven possible states
according to the remaining players. These are labeled as
$0,1,\dots,6$. There are transitions between those states, as
shown in the diagram in Fig.~\ref{fig:random}, where $p_{ij}$
denotes the transition probability from state $i$ to state $j$
(the self--transition probability $p_{ii}$ is denoted by $r_i$).

From Markov chain theory\cite{k75.2} we can evaluate the
probability $u^j_i$ that starting from state $i$ we eventually end
up in state $j$ after a sufficiently large number of steps. In
particular, if we start from state $0$ (with the three players
active), the nature of the game is such that the only
non-vanishing probabilities are $u_0^4$, $u_0^5$ and $u_0^6$
corresponding to the winning of the game by player A, B and C
respectively. The relevant set of equations is~\footnote{There is
no need to write down the equations for $u_0^6$ since it suffices
to notice that $u_0^4+u_0^5+u_0^6=1$.}:

\begin{equation}
\begin{array}{lll}
&u_0^4= p_{01}~u_1^4+p_{02}~u_2^4+p_{03}~u_3^4+r_0~ u_0^4,
&u_0^5= p_{01}~u_1^5+p_{02}~u_2^5+p_{03}~u_3^5+r_0~ u_0^5,\nonumber\\
&u_1^4 = p_{14}~u_4^4+r_1~ u_1^4,
&u_1^5 = p_{15}~u_5^5+r_1~ u_1^5,\nonumber\\
&u_2^4 = p_{24}~u_4^4+r_2~ u_2^4,
&u_2^5 = r_2~u_2^5,\\
&u_3^4 = r_3~ u_3^4, &u_3^5 = r_3~ u_3^5+p_{35}~u_5^5.\nonumber
\end{array}
\end{equation}

Solving for $u_0^4$, $u_0^5$ and $u_0^6$ we obtain:

\begin{eqnarray}
&u_0^4& = \frac{p_{01}~p_{14}}{(1-r_0)(1-r_1)} + \frac{p_{02}~p_{24}}{(1-r_0)(1-r_2)}\label{u04},\nonumber\\
&u_0^5& = \frac{p_{01}~p_{15}}{(1-r_0)(1-r_1)} + \frac{p_{03}~p_{35}}{(1-r_0)(1-r_3)}\label{u05},\\
&u_0^6& = \frac{p_{02}~p_{26}}{(1-r_0)(1-r_2)} +
\frac{p_{03}~p_{36}}{(1-r_0)(1-r_3)}.\label{u06}\nonumber
\end{eqnarray}

We can now derive the expressions for the transition probabilities
$p_{ij}$. Remember that we denote by $a$ the probability that
player A eliminates from the game the player he has aimed at (and
similarly for $b$ and $c$), and that $P_{\alpha\beta}$
($\alpha=$A,C,B and $\beta=$ A,B,C,0) the probability of player $\alpha$
choosing player $\beta$ (or into the air if $\beta=0$) as a target
when it is his turn to play (a situation that only appears when
the three players are still active). We have then:


\begin{equation}
\begin{array}{lll}
&r_{0} = 1-\frac{1}{3}(a(1-P_{A0})+b(1-P_{B0})+c(1-P_{C0})),&p_{01} = \frac{1}{3}(aP_{AC}+bP_{BC}),\\ &p_{02}=\frac{1}{3}(aP_{AB}+cP_{CB}),&p_{03} = \frac{1}{3}(bP_{BA}+cP_{CA}),\\
&p_{14} = p_{24} = \frac{1}{2}a, &p_{15} = p_{35} = \frac{1}{2}b,\\
&p_{26} = p_{36} = \frac{1}{2}c,& r_1 = 1-\frac{1}{2}(a+b),\\
&r_2= 1-\frac{1}{2}(a+c),&r_3 = 1-\frac{1}{2}(b+c).
\end{array}
\end{equation}

\subsection{Sequential firing}
\label{ap:seq}

As in the random firing case, we describe this game as a Markov
chain composed of $11$ different states, also with three absorbent
states: $9$ , $10$ and $11$. In Fig.~\ref{fig:seq} we can see the
corresponding diagram for this game, together with a table
describing all possible states. Based on this diagram, we can
write down the relevant set of equations for the transition
probabilities $u_i^j$:

\begin{figure}
\begin{tabular}{|c|c|}
\hline
\textit{States} & \textit{Remaining players}\\
\hline
$0$ & A B \textbf{C} \\
$1$ & A \textbf{B} C\\
$2$ & \textbf{A} B C\\
$3$ & \textbf{B} C \\
$4$ & \textbf{A} C\\
$5$ & B \textbf{C}\\
$6$ & \textbf{A} B\\
$7$ & A \textbf{C}\\
$8$ & A \textbf{B}\\
$9$ & C\\
$10$ & B\\
$11$ & A\\
\hline
\end{tabular}
\epsfig{figure=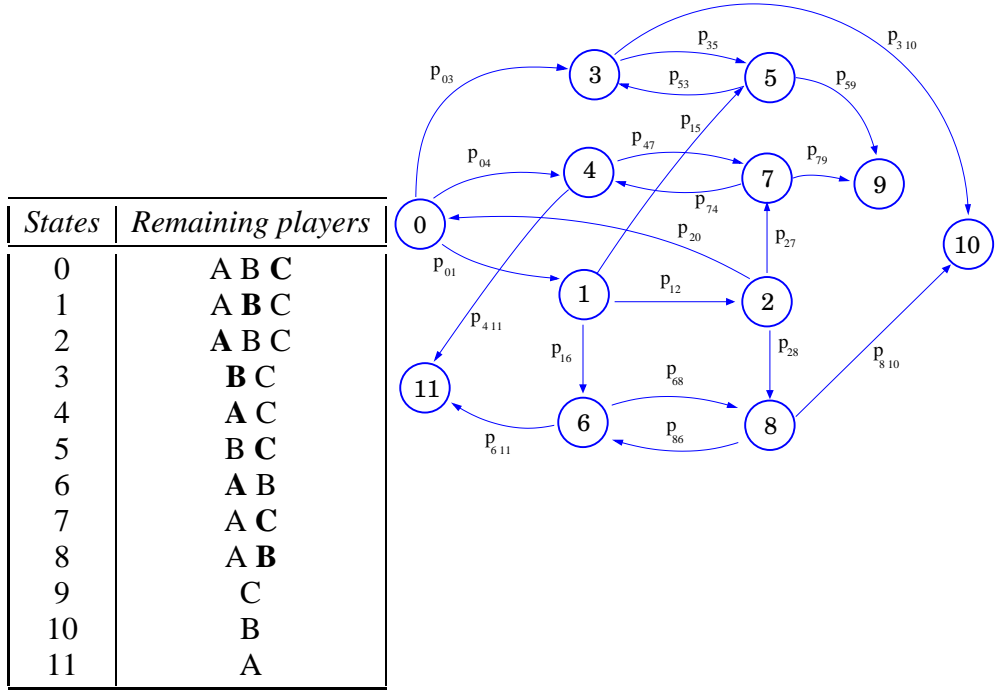,width=8.cm}
\caption{\label{fig:seq}Table: Description of the different states
of the game for the case of sequential firing. The highlighted
player is the one chosen for shooting in that state. Diagram:
scheme representing all the allowed transitions between the states
shown in the table for the case of a truel with sequential firing in
the order C$\to$ B $\to$ A with $a>b>c$.}
\end{figure}

\begin{equation}
\begin{array}{llll}
\vspace{0.2cm}&u_0^9 = p_{03}u_3^9+p_{01}u_1^9+p_{04}u_4^9,
&u_0^{10} =p_{03}u_3^{10}+p_{01}u_1^{10},
& u_0^{11} = p_{01}u_1^{11}+p_{04}u_4^{11},\\
\vspace{0.2cm}&
u_1^{10}=p_{12}u_2^{10}+p_{15}u_5^{10}+p_{16}u_6^{10}, & u_1^9
=p_{12}u_2^9+p_{15}u_5^9,
& u_1^{11}= p_{12}u_2^{11}+p_{16}u_6^{11},\\
\vspace{0.2cm}&
u_2^{11}=p_{28}u_8^{11}+p_{27}u_7^{11}+p_{20}u_0^{11}, & u_2^9
=p_{27}u_7^9+p_{20}u_0^9,
&u_2^{10}=p_{28}u_8^{10}+p_{20}u_0^{10},\\
\vspace{0.2cm}&u_3^9 = p_{35}u_5^9,
&u_3^{10}= p_{35}u_5^{10}+p_{3\,10},\\
\vspace{0.2cm}&u_4^9 = p_{47}u_7^9,
&u_4^{11} = p_{47}u_7^{11}+p_{4\,11},\\
\vspace{0.2cm}&u_5^9 = p_{53}u_3^9+p_{59},
&u_5^{10} = p_{53}u_3^{10},\\
\vspace{0.2cm}&u_6^{10} = p_{68}u_8^{10},
&u_6^{11} = p_{68}u_8^{11}+p_{6\,11},\\
\vspace{0.2cm}&u_7^9 = p_{74}u_4^9+p_{79},
&u_7^{11} =p_{74}u_4^{11},\\
\vspace{0.2cm}&u_8^{10} = p_{86}u_6^{10}+p_{8\,10}, &u_8^{11} =
p_{86}u_6^{11}.
\end{array}
\end{equation}

The general solutions for the probabilities $u_0^9$, $u_0^{10}$
and $u_0^{11}$ are given by
\begin{eqnarray}
u_0^9&=&\frac{1}{1-p_{01}p_{12}p_{20}}\left[\frac{p_{59}(
p_{03}p_{35} + p_{01}p_{15})}{1-p_{35}p_{53}} +
\frac{p_{79}(p_{04}p_{47}+p_{01}p_{12}p_{27})}{1-p_{47}p_{74}}\right],
\nonumber\\
u_0^{10}&=&\frac{1}{1-p_{01}p_{12}p_{20}}\left[\frac{p_{3\,10}(
p_{03} + p_{01}p_{15}p_{53})}{1-p_{35}p_{53}} +
\frac{p_{01}p_{8\,10}(p_{16}p_{68}+
p_{12}p_{28})}{1-p_{68}p_{86}}\right],
\\
u_0^{11}&=&\frac{1}{1-p_{01}p_{12}p_{20}}\left[\frac{p_{4\,11}(
p_{04} + p_{01}p_{12}p_{27}p_{74})}{1-p_{47}p_{74}} +
\frac{p_{01}p_{6\,11} ( p_{16}+ p_{12}p_{28}p_{86})}
{1-p_{68}p_{86}}\right], \nonumber
\end{eqnarray}

with transition probabilities given by

\begin{equation}
\begin{array}{llll}
&p_{01}=(1-c)+c P_{C0}, &p_{03}=cP_{CA}, &p_{04}=cP_{CB},\nonumber\\
&p_{12}=(1-b)+b P_{B0}, &p_{15}=bP_{BA}, &p_{16}=bP_{CA},\nonumber\\
&p_{20}=(1-a)+a P_{A0}, &p_{27}=aP_{AB}, &p_{28}=aP_{AC},\nonumber\\
&p_{35}=p_{86} = 1-b, &p_{3\,10}=p_{8\,10} = b,\nonumber\\
&p_{47}=p_{68} = 1-a, &p_{4\,11}=p_{6\,11} = a,\nonumber\\
&p_{53}=p_{74} = 1-c, &p_{59}=p_{79} = c.
\end{array}
\end{equation}

\subsection{Convincing opinion}
\label{ap:con} 
\begin{figure}[!htb]
\begin{tabular}{|c|c|}
\hline
\textit{States} & \textit{Opinions}\\
\hline
$0$ & A B C\\
$1$ & A A A\\
$2$ & B B B\\
$3$ & C C C\\
$4$ & A A B\\
$5$ & A B B\\
$6$ & A A C\\
$7$ & A C C\\
$8$ & B B C\\
$9$ & B C C\\
\hline
\end{tabular}
\epsfig{figure=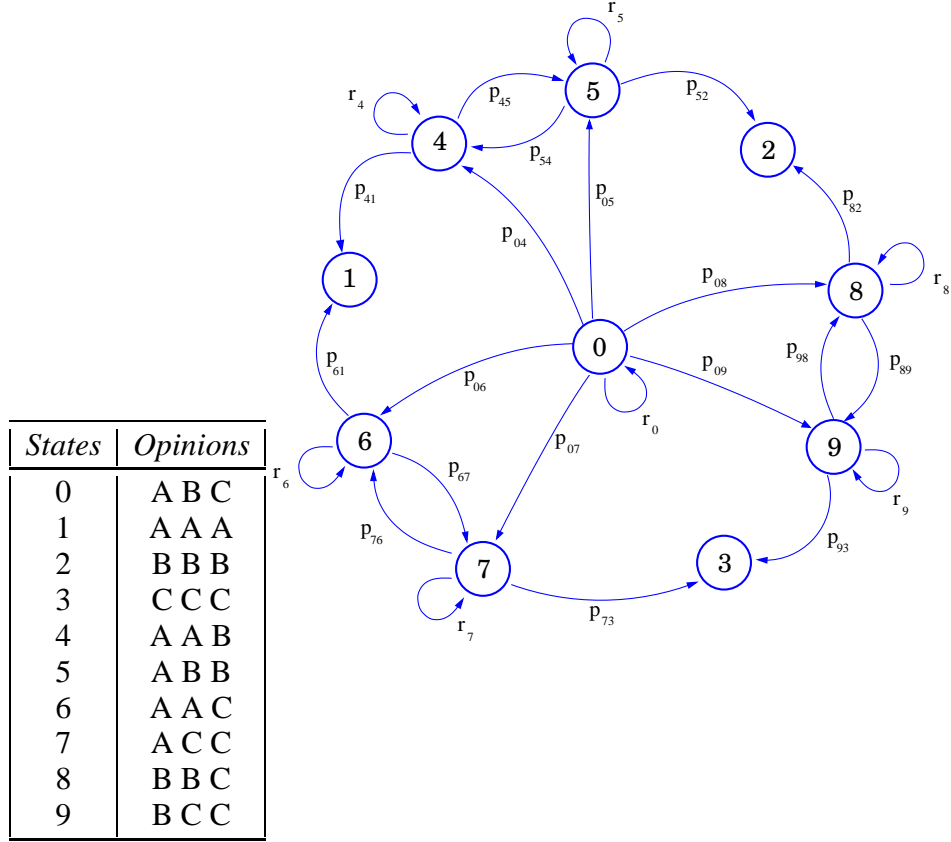,width=9.cm}
\caption{\label{fig:opinion}Table: description of the different
states of the opinion model. Diagram: scheme representing the
allowed transitions between the states. }
\end{figure}

For this model we show in  Fig.~\ref{fig:opinion}
the diagram of all the allowed states and transitions, together
with a table describing the possible states.

The corresponding set of equations describing this convincing
opinion model, as derived from the diagram, are
\begin{equation}
\begin{array}{lll}
&u_0^1= r_0u_0^1+p_{06}u_6^1+p_{04}u_4^1+p_{05}u_5^1+p_{07}u_7^1,\\
&u_0^2= r_0u_0^2+p_{04}u_4^2+p_{05}u_5^2+p_{08}u_8^2+p_{09}u_9^2,\\
&u_0^3= r_0u_0^3+p_{08}u_8^3+p_{09}u_9^3+p_{07}u_7^3+p_{06}u_6^3,\\
&u_4^1= r_4u_4^1+p_{45}u_5^1+p_{41},
&u_4^2= r_4u_4^2+p_{45}u_5^2,\\
&u_5^1= r_5u_5^1+p_{54}u_4^1,
&u_5^2= r_5u_5^2+p_{54}u_4^2+p_{52},\\
&u_6^1= r_6u_6^1+p_{67}u_7^1+p_{61},
&u_6^3= r_6u_6^3+p_{67}u_7^3,\\
&u_7^1= r_7u_7^1+p_{76}u_6^1,
&u_7^3= r_7u_7^3+p_{76}u_6^3+p_{73},\\
&u_8^2= r_8u_8^2+p_{89}u_9^2+p_{82},
&u_8^3= r_8u_8^3+p_{89}u_9^3,\\
&u_9^2= r_9u_9^2+p_{98}u_8^2, &u_9^3= r_9u_9^3+p_{98}u_8^3+p_{93}.
\end{array}
\end{equation}

And the general solution for the probabilities $u_0^1$, $u_0^2$
and $u_0^3$ is
\begin{eqnarray}
u_0^1&=&\frac{1}{1-r_0}\left[\frac{p_{61}(
p_{06}(1-r_7)+p_{07}p_{76}
)}{(1-r_6)(1-r_7)-p_{67}p_{76}}+\frac{p_{41}(p_{04} ( 1-r_5) +
p_{05}p_{54})}{(1-r_4)(1-r_5)-p_{45}p_{54}}\right],\nonumber\\
u_0^2&=&\frac{1}{1-r_0}\left[\frac{p_{52}(
p_{04}p_{45}+p_{05}(1-r_4)
)}{(1-r_4)(1-r_5)-p_{45}p_{54}}+\frac{p_{82}(p_{08}( 1-r_9) +
p_{09}p_{98})}{(1-r_8)(1-r_9)-p_{89}p_{98}}\right],\nonumber\\
u_0^3&=&\frac{1}{1-r_0}\left[\frac{p_{73}(
p_{06}p_{67}+p_{07}(1-r_6)
)}{(1-r_6)(1-r_7)-p_{67}p_{76}}+\frac{p_{93}(p_{09}( 1-r_8) +
p_{08}p_{89})}{(1-r_8)(1-r_9)-p_{89}p_{98}}\right],
\end{eqnarray}
where the transition probabilities are given by
\begin{equation}
\begin{array}{llll}
&p_{04}=\frac{1}{3}cP_{CA}, &p_{06}=\frac{1}{3}cP_{CB},
&p_{08}=\frac{1}{3}bP_{BC},\\
&p_{05}=\frac{1}{3}bP_{BA}, &p_{07}=\frac{1}{3}aP_{AB},
&p_{09}=\frac{1}{3}aP_{AC},\\
&p_{41}= p_{61} = \frac{2}{3}c &p_{45}=p_{98} = \frac{1}{3}b,
&p_{54}= p_{76} = \frac{1}{3}c,\\
&p_{52}= p_{82} = \frac{2}{3}b, &p_{67}= p_{89} = \frac{1}{3}a,
&p_{73}= p_{93} = \frac{2}{3}a,\\
&r_0=\frac{1}{3}[3-a-b-c], &r_4=\frac{2}{3}(1-c)+\frac{1}{3}(1-b),
&r_5=\frac{1}{3}(1-c)+\frac{2}{3}(1-b),\\
&r_6=\frac{2}{3}(1-c)+\frac{1}{3}(1-a),
&r_7=\frac{1}{3}(1-c)+\frac{2}{3}(1-a),
&r_8=\frac{2}{3}(1-b)+\frac{1}{3}(1-a),\\
&r_9=\frac{1}{3}(1-b)+\frac{2}{3}(1-a).&&
\end{array}
\end{equation}

\begin{theacknowledgments}
{\bf Acknowledgments} We thank Ces\'areo Hern\'andez for bringing
this problem to our attention. This work is supported
by MCyT (Spain) and FEDER (EU) projects FIS2004-5073-C04-03 and
FIS2004-953; P.A. acknowledges support form the Govern Balear,
Spain.
\end{theacknowledgments}

\end{document}